\documentclass[]{emulateapj}

\usepackage{natbib}
\usepackage{lineno}
\usepackage{multirow}
\usepackage{booktabs}
\usepackage{epsfig}
\usepackage{xcolor}
\usepackage{ulem,hyperref}
\usepackage{amsmath}
\usepackage{soul} 
\newcommand{\wCDM}{$w$CDM}
\newcommand{\wwCDM}{$w_0w_a$CDM}
\newcommand{\LCDM}{$\Lambda$CDM}

\newcommand{\PBIa}{{\cal P}_{\rm B(Ia)} }
\newcommand{\PIa}{P_{\rm Ia}}
\newcommand{\DIa}{D_{\rm Ia}}
\newcommand{\DCC}{D_{\rm CC}}

\newcommand{\PBIai}{{\cal P}_{{\rm B(Ia)},i} }
\newcommand{\PIai}{P_{{\rm Ia},i}}
\newcommand{\DIai}{D_{{\rm Ia},i}}
\newcommand{\DCCi}{D_{{\rm CC},i}}

\newcommand{\OM}{\rm \Omega_{\rm M}}
\newcommand{\OL}{\rm \Omega_\Lambda}

\newcommand{\mubias}{\Delta\mu_{\rm bias}}
\newcommand{\muref}{\mu_{\rm ref}}

\newcommand{\M}{{\cal M}}
\newcommand{\C}{\vec{\cal C}}
\newcommand{\Cref}{\vec{\cal C}_{\rm ref}}

\newcommand{\HR}{{\rm HR}}

\newcommand{\HRavg}{\langle{\rm HR}_{\zeta}\rangle}

\newcommand{\LK}{{\cal L}}
\newcommand{\chiHR}{\chi^2_{\rm HR}}
\newcommand{\sigmu}{\sigma_{\mu}}
\newcommand{\sigmui}{\sigma_{\mu,i}}

\newcommand{\sigunbini}{\sigma_{\mu,{\rm unbin},i}}

\newcommand{\zbin}{z_{\zeta}}
\newcommand{\Mz}{M_{\zeta}}
\newcommand{\mubin}{\mu_{\zeta}}
\newcommand{\sigM}{\sigma_{M,\zeta}}
\newcommand{\Sbin}{S_{\zeta}}
\newcommand{\zxc}{\vec{\zeta}}

\newcommand{\sigint}{\sigma_{\rm int}}
\newcommand{\mutrue}{\mu_{\rm true}}
\newcommand{\wbias}{\Delta w}

\newcommand{\sigwavg}{$\langle\sigma_w\rangle$}
\newcommand{\sigwoavg}{$\langle\sigma_{w0}\rangle$}
\newcommand{\sigwaavg}{$\langle\sigma_{wa}\rangle$}
\newcommand{\wbiasavg}{${\langle}w$-bias${\rangle}$}
\newcommand{\wobiasavg}{${\langle}w_0$-bias${\rangle}$}
\newcommand{\wabiasavg}{${\langle}w_a$-bias${\rangle}$}
\newcommand{\Nsigma}{$N_{\sigma}$}
\newcommand{\FoMavg}{$\langle${FoM}$\rangle$}

\newcommand{\SNANA}{{\tt SNANA}}

\newcommand{\MVinprep}{M.~Vincenzi et al. in preparation}
\newcommand{\LOWZ}{LOWZ}

\newcommand{\URLSNANA}{\url{https://github.com/RickKessler/SNANA}}
\newcommand{\URLSNN}{\url{https://github.com/supernnova}}
\newcommand{\URLPIPPIN}{\url{https://github.com/dessn/Pippin}}

\newcommand{\URLPS}{\url{https://panstarrs.stsci.edu}}
\newcommand{\URLDES}{\url{https://www.darkenergysurvey.org}}
\newcommand{\URLSDSS}{\url{https://www.sdss.org/dr12/data_access/supernovae}}
\newcommand{\URLLSST}{\url{https://www.lsst.org}}
\newcommand{\URLRoman}{\url{https://roman.gsfc.nasa.gov}}
\newcommand{\URLIax}{\url{https://github.com/RutgersSN/SNIax-PLAsTiCC}}

\newcommand{\spec}{spectroscopic}
\newcommand{\specy}{spectroscopically}
\newcommand{\obs}{observation}

\newcommand{\sigR}{0.006}  % sigma_R for R-shift parameter used in CMB prior
\newcommand{\NSAMPLE}{50}  % number of simulated samples
\newcommand{\OMref}{0.3}
\newcommand{\wref}{-1}

\newcommand{\nbz}{20}
\newcommand{\nbxA}{2} % number of stretch bins for rebinA
\newcommand{\nbcA}{4} % number of color bins for rebinA
\newcommand{\nbxB}{4} % number of stretch bins for rebinB
\newcommand{\nbcB}{8} % number of color bins for rebinB

\newcommand{\nbtotA}{160}   % total number of HD bins for Rebin2x4
\newcommand{\nbtotB}{640}   % total number of HD bins for Rebin4x8

\newcommand{\RebinA}{Rebin2x4}
\newcommand{\RebinB}{Rebin4x8}

\newcommand{\NSNavgTOT}{1897}
\newcommand{\NSNavgDES}{1622}
\newcommand{\NSNavgFOUND}{172}
\newcommand{\NSNavgLOWZ}{103}

\newcommand{\RebinFNA}{Rebinning with \nbxA\ stretch bins and \nbcA\ color bins}
\newcommand{\RebinFNB}{Rebinning with \nbxB\ stretch bins and \nbcB\ color bins}
\newcommand{\NsigwBiasFN}{Absolute value of \wbiasavg\ divided by its uncertainty}
\newcommand{\NsigwoBiasFN}{Absolute value of \wobiasavg\ divided by its uncertainty}
\newcommand{\NsigwaBiasFN}{Absolute value of \wabiasavg\ divided by its uncertainty}

\newcommand{\Sbinavg}{1.01}
\newcommand{\NBAND}{34}
\newcommand{\NSYST}{70} % 34 zp + 34 wave + HST + MWEBBV

\newcommand{\FoMbin}{45}
\newcommand{\FoMunbin}{55}    % (56 -> 55, May 2 2023)
\newcommand{\FoMrebinA}{51}   % rebin2x4
\newcommand{\FoMrebinB}{54}   % rebin4x8  (55->54, May 2023)

\begin{document}

\title{Binning is Sinning: Redemption for Hubble Diagram
 Using \\ Photometrically Classified Type Ia Supernovae }

\def\andname{}

\author{
R.~Kessler\altaffilmark{1,2},
M.~Vincenzi\altaffilmark{3},
P.~Armstrong\altaffilmark{4}
}

\affil{$^{1}$ Kavli Institute for Cosmological Physics, University of Chicago, Chicago, IL 60637, USA}
\affil{$^{2}$ Department of Astronomy and Astrophysics, University of Chicago, Chicago, IL 60637, USA}
\affil{$^{3}$ Department of Physics, Duke University, Durham, NC 27708, USA}
\affil{$^{4}$ Mt Stromlo Observatory, The Research School of Astronomy and Astrophysics,   
Australian National University, Stromlo, ACT 2601, Australia}

%\vspace{3in}

\begin{abstract}
Bayesian Estimation Applied to Multiple Species (BEAMS) is implemented
in the BEAMS with Bias Corrections (BBC) framework to
produce a redshift-binned Hubble diagram (HD) 
for Type Ia supernovae (SNe~Ia).
BBC corrects for selection effects and non-SNIa contamination,
and systematic uncertainties are described by a covariance matrix with 
dimension matching the number of BBC redshift bins.
For \specy\ confirmed SN~Ia samples, a recent ``Binning is Sinning" article 
(BHS21, arxiv:2012.05900)
showed that an unbinned HD and covariance matrix reduces the systematic uncertainty by 
a factor of ${\sim}1.5$ compared to the binned approach.
Here we extend their analysis to obtain an unbinned HD for a
photometrically identified sample processed with BBC.
To test this new method, 
we simulate and analyze \NSAMPLE\ samples
corresponding to the Dark Energy Survey (DES)
with a low-redshift anchor;
the simulation includes SNe~Ia, and contaminants from core-collapse SNe and peculiar SNe~Ia.
The analysis includes systematic uncertainties
for calibration,
and measures the dark energy equation of state parameter ($w$).
Compared to a redshift-binned HD, the unbinned HD 
with nearly 2000 events results in a smaller 
systematic uncertainty, in qualitative agreement with BHS21,
and averaging results among the \NSAMPLE\ samples 
we find no evidence for a $w$-bias.
To reduce computation time for fitting an unbinned HD with large samples,
we propose an HD-rebinning method that defines the HD
in bins of redshift, color, and stretch; the rebinned HD results in similar uncertainty
as the unbinned case, and shows no evidence for a $w$-bias.
\end{abstract}


% ------------------------------------
\section{Introduction}
\label{sec:intro} 

Following the discovery of cosmic acceleration using a few dozen
Type Ia supernovae (SNe~Ia) \citep{Riess98,Perlmutter99}, 
increasingly large SN~Ia samples have been used to improve measurements of the
dark energy equation of state parameter, $w$.
While the most precise $w$ measurements are based on \specy\ confirmed
samples, photometric imaging surveys have been discovering far more supernovae than 
\spec\ resources can observe.
Existing SN surveys include the 
Sloan Digital Sky Survey-II (SDSS),\footnote{\URLSDSS}
Supernova Legacy Survey (SNLS),
Panoramic Survey Telescope and Rapid Response System-1 (PS1),\footnote{\URLPS} and
Dark Energy Survey (DES);\footnote{\URLDES} 
future wide-area surveys that will overwhelm \spec\ resources
include the Legacy Survey of Space and Time (LSST)\footnote{\URLLSST} and the
Nancy Grace Roman Space Telescope.\footnote{\URLRoman}

To make full use of these large SN~Ia samples, photometric identification
using broadband filters has been developed over the past decade.
Photometric methods include template matching \citep{Sako2011} 
and machine learning \citep{Lochner2016,Moller2020,Qu2021},
and they determine the probability ($\PIa$) for each event to be an SN~Ia.
A framework to incorporate the resulting $\PIa$  was
developed to measure cosmological parameters.
This framework, called ``Bayesian Estimation Applied to Multiple Species"
(BEAMS: \citet{BEAMS2007,BEAMS2012}),
was first used in a SN-cosmology analysis for the PS1 photometric sample
\citep{Jones2018}.

As part of the DES SN-cosmology analysis, BEAMS was extended to
``BEAMS with Bias Corrections" (BBC: \citet{bbc}; hereafter KS17), a fitting procedure 
designed to produce a Hubble diagram (HD) that is corrected for selection biases
and for contamination from core-collapse SNe (SNCC) and peculiar SNe~Ia.
BBC has been used in the SN Ia cosmology analysis for \spec\ samples 
from Pantheon \citep{Pantheon}, DES \citep{des3yr}, and Pantheon+ \citep{PantheonPlus}.
BBC has also been used on a photometric sample from PS1 \citep{Jones2019},
and to examine contamination biases for the 
photometric DES sample \citep{Vinc2023_bias}.

The BBC fit is performed in redshift bins to determine nuisance parameters 
and SN~Ia distances that are independent of cosmological parameters,
which enables more flexible use of cosmology-fitting programs.
BBC therefore produces both a redshift binned and unbinned HD.
Previous analyses using BBC took advantage of the binned HD
to reduce computation time in cosmology-fitting programs.
However, \citet{Binning_is_Sinning} (hereafter BHS21),
showed that while the statistical uncertainty is the same using a
binned or unbinned HD,
the systematic uncertainty is ${\sim}1.5$ smaller using an unbinned 
HD and unbinned covariance matrix. 
The uncertainty reduction is from an effect known as self-calibration
\citep{SelfCalib2011}.

BHS21 demonstrated the uncertainty reduction using BBC with a \specy\ 
confirmed sample. Here we expand the use of unbinned HDs
to photometric samples where BEAMS is used. 
In anticipation of very large samples in future analyses, we also explore the possibility 
of reducing computation time with a smaller HD and covariance matrix and still benefit 
from self-calibration: 
a rebinned HD in the space of redshift, color, and stretch.
This choice of variables is motivated by the color-dependent systematic explored in BHS21.
While the unbinned approach is optimal, the rebinned approach may be useful
for the many intermediate simulation tests prior to unblinding.

We validate the unbinning and rebinning methods using simulations of DES that include
SNe~Ia, SNCC, and peculiar SNe~Ia. The simulation and analysis presented
here are similar to those in \citet{Vinc2023_bias},
and all analysis software used in this analysis is publicly available.
The software for simulations, light-curve fitting, BBC,
and cosmology fitting is from the 
\textbf{S}uper\textbf{N}ova \textbf{ANA}lysis package 
({\SNANA}; \citet{Kessler09SNANA}).\footnote{\URLSNANA}
The photometric classification software is from
{\tt SuperNNova} (SNN; \citet{Moller2020}).\footnote{\URLSNN}
For workflow orchestration we used {\tt Pippin} \citep{Pippin}.\footnote{\URLPIPPIN}

The outline of this Letter is as follows.
The SALT2 and BBC formalism is reviewed in \S\ref{sec:review}.
The unbinning and rebinning procedures are presented in \S\ref{sec:unbin}.
The validation analysis is described in \S\ref{sec:validate1},
and the validation results are given in \S\ref{sec:validate2}.

% =================================================================================
\section{Review of SALT2 and BBC}
\label{sec:review}

Using the SALT2 framework \citep{Guy2010}, 
BBC is a fitting procedure that delivers an HD corrected 
for selection effects and corrected for contamination. BBC incorporates
three main features:
(1) BEAMS \citep{BEAMS2007}, 
(2) fitting in redshift bins to avoid dependence on cosmological parameters \citep{Marriner2011}, and
(3) detailed simulation to correct distance biases \citep{Kessler2019_sim}.

For each event, a SALT2 light-curve fit determines the 
time of peak brightness ($t_0$),
stretch ($x_1$), 
color parameter ($c$), and
amplitude ($x_0$) with $m_B\equiv -2.5\log_{10}(x_0)$.
Within BBC, the measured distance modulus is defined by the 
Tripp equation \citep{Tripp1998}:
\begin{equation}
   \mu = m_B + \alpha x_1 - \beta c + \M - \mubias~,  
   \label{eq:mu}
\end{equation}
where $\alpha$ and $\beta$ are the stretch- and color-luminosity parameters,
$\M$ is a global offset, and
$\mubias$ is a distance-bias correction for each event, $\mu - \mutrue$,
determined from a large simulation. The $\mubias$ value for each event is evaluated
by interpolating in a 5-dimensional space of $\{z,x_1,c,\alpha,\beta\}$.
In Eq.~\ref{eq:mu}, there is an implicit SN index $i$ for 
$\mu,m_B,x_1,c$, and $\mubias$;
this index is suppressed for readability.
To simplify this study,
host-SN correlations have been ignored in Eq.~\ref{eq:mu}, 
and also in the simulations used for validation.

Following Section~5 of KS17 and making a few simplifications for this review, 
the BBC fit maximizes a likelihood of the form $\LK = \prod_{i=1}^N\LK_i$,
where $\LK_i$ for event $i$ is
\begin{equation}
    {\LK}_i = \PIai \DIai + (1-\PIai)\DCCi~,
\end{equation}
where $\PIai$ is the photometric classification probability for 
event $i$ to be an SN~Ia.
The SN~Ia component of $\LK_i$ is $\DIa{\sim}\exp[-\chiHR/2]$, 
where $\chiHR = \HR^2/\sigmu^2$, 
$\HR$ is a Hubble residual described below, and
$\sigmu$ is the uncertainty on $\mu$ in Eq.~\ref{eq:mu} as shown in Eq.~3 of KS17.
The non-SNIa (contamination) component, $\DCC$, 
is evaluated from a simulation.

To remove the dependence on cosmological parameters in the BBC fit, 
we follow \citet{Marriner2011} and define the Hubble residual for 
the $i$'th SN ($\HR_i$) as
\begin{equation}
   \HR_i \equiv \mu_i - [ \muref(z_i,\Cref) + \Mz ]
   \label{eq:HR}
\end{equation}
where $\muref = \muref(z_i,\Cref)$ is a reference distance computed from redshift $z_i$ and
an arbitrary choice of reference cosmology parameters denoted by $\Cref$. 
Our choice for $\Cref$ is flatness and
\begin{equation}
    \Cref \equiv \{\OM=\OMref, w=\wref\}~.
    \label{eq:Cref}
\end{equation}
$\Mz$ are fitted distance offsets in redshift bins denoted by $\zeta$.
An important concept is that using a cosmological model for $\muref$ in Eq.~\ref{eq:HR} 
is a convenience, not a necessity. 
For example, $\muref$ could be replaced with a polynomial function of redshift
or any function that approximates the distance-redshift relation within each redshift bin.

The BBC fit determines $\alpha$, $\beta$, $\gamma$, $\Mz$, and an intrinsic scatter
term ($\sigint$) added to the distance uncertainties (\S\ref{sec:unbin})
that results in a reduced $\chi^2$ of one. 
The final binned HD is obtained as follows.
First, each binned redshift ($\zbin$) is computed from
\begin{equation}
    \zbin = \mu^{-1}[\overline{\muref}_{\zeta}]
\end{equation}
where $\mu^{-1}$ is an inverse-distance function that numerically determines 
redshift from the weighted average of $\muref$ in redshift bin $\zeta$.
The weight for each event is $\sigmu^{-2}$.
Next, the BBC-fitted distance in each redshift bin ($\mubin$) is
\begin{equation}
    \mubin = \overline{\muref}_{\zeta} + \Mz. 
\end{equation}
and the collection of $\{\zbin, \mubin\}$ is the binned HD
corrected for selection effects and contamination. 
The uncertainty on $\mubin$ is the BBC-fitted uncertainty for $\Mz$.
If a different choice of $\C$ is used for $\muref$,  
the fitted $\Mz$ will change but the $\mubin$ remain the same.
For \specy\ confirmed samples with all ${\PIa}_{,i}=1$,
the unbinned HD is the collection of $\{z_i, \mu_i\}$
where the $\mu_i$ are computed from Eq.~\ref{eq:mu} using the 
BBC-fitted parameters 
and each distance uncertainty ($\sigmui$)
is computed from Eq.~3 in KS17.
This procedure is an approximation that
we rigorously test (\S\ref{subsec:Cref}) with high-statistics simulations.

% ==================================================================
\section{Unbinning and Rebinning after BBC Fit}
\label{sec:unbin}

For an unbinned HD, we use the BBC-fitted parameters and
compute the distances defined in Eq.~\ref{eq:mu}. 
The unbinned distance uncertainties ($\sigunbini$), however, are not 
the naively computed distance uncertainties 
($\sigmui$) for a \specy\ confirmed sample.
To determine $\sigunbini$, we require that the weighted average uncertainty
in each redshift bin is equal to $\sigM$, the BBC-fitted uncertainty on $\Mz$;
\begin{eqnarray}
    1/\sigM^2 
      & = & \sum_{i\in \zeta} 1/\sigunbini^2           \label{eq:sigM} \\ 
      & = & \sum_{i\in \zeta} \PBIai/[\Sbin\sigmui]^2  \label{eq:sigM2}
\end{eqnarray}
where $i$ is the SN index within redshift bin $\zeta$,
\begin{equation}
   \PBIai =  \frac{ \PIai \DIai }{ \PIai\DIai + (1-\PIai)\DCCi } \label{eq:PBIa}
\end{equation}
is the BEAMS probability for event $i$ to be an SN~Ia, and
$\Sbin$ is a $\zeta$-dependent uncertainty scale that is computed to satisfy Eq.~\ref{eq:sigM2}.
We find that $\Sbin$ is a few percent greater than 1 because of small correlations
between the fitted parameters ($\alpha$,$\beta$,$\Mz$).
Eq.~\ref{eq:sigM2} is an ad hoc assumption and does not have a rigorous derivation.
From Eqs.~\ref{eq:sigM}-\ref{eq:sigM2}, the unbinned distance uncertainty is
\begin{equation}
    \sigunbini = \Sbin\sigmui \Big/ \sqrt{\PBIai}~.   
    \label{eq:sigunbin} 
\end{equation}

As a crosscheck, the weighted average of the Hubble residuals ($\HRavg$)
should be zero for each redshift bin:
\begin{equation}
    \HRavg = \Big[ \sum_{i \in \zeta} \HR_i W_i \Big] \Big/ 
        \Big[ \sum_{i \in \zeta} W_i \Big] 
        \label{eq:HRavg}
\end{equation}
where $W_i = \sigunbini^{-2}$
and $\HR_i$ is defined in Eq.~\ref{eq:HR}.

% ==================================================================
%\section{Rebinning with BBC}

To limit the size of the HD and still benefit from reduced systematics,
we propose rebinning in the space of redshift, stretch, and color, 
denoted by $\zxc = \{z,x_1,c\}$.
The distance modulus in each 3D $\zxc$ cell is a weighted average of distances in 
the cell,
\begin{equation}
    \mu_{\zxc} =  \sum_{i \in\zxc}\mu_i W_i \Big/ \sum_{i\in\zxc} W_i
\end{equation}
and following Eq.~\ref{eq:sigM} the uncertainty on $\mu_{\zxc}$ is
\begin{equation}
     1/\sigma^2_{\mu,\zxc} =  \sum_{i\in \zxc} 1/\sigunbini^2
    \label{eq:sigzxc}
\end{equation}

% ==================================================================
\section{Validation I: Simulation and Analysis}
\label{sec:validate1}

\newcommand{\URLCSP}{\url{https://csp.obs.carnegiescience.edu}}

We test the unbinning and rebinning procedure (\S\ref{sec:unbin})
by analyzing $\NSAMPLE$ simulated data-sized samples that closely follow
\citet{Vinc2023_bias}.
Each simulation corresponds to the 5-year DES photometric sample for events
with an accurate \spec\ redshift of the host galaxy, 
combined with a \specy\ confirmed low-redshift (\LOWZ) sample ($z<0.1$).
The \LOWZ\ sample uses the cadence and 
signal-to-noise ratio (S/N) properties for the 
Carnegie Supernova Project (CSP),\footnote{\URLCSP}
Center for Astrophysics (CFA3,CFA4; \citet{Hicken2009_CfA3,Hicken2012}),
and Foundation Supernova Survey \citep{Found2018}.

The simulated models include:  
\begin{itemize}
    \setlength\itemsep{0.01em}
    \item SNe~Ia generated from the SALT2 model in \citet{Guy2010}
      using trained model parameters from \citet{Betoule14};
    \item SNCC generated from spectral energy distribution (SED) 
            templates in \citet{Vinc2019};
    \item Peculiar SN Iax using the SED model from \citet{PLASTICC}\footnote{\URLIax} 
         and extinction correction from \citet{Vinc2021_sim};
    \item Peculiar 91bg-like SNe~Ia using the SED model from \citet{PLASTICC}; and
    \item \LCDM\  with $\OM=0.311$, $\OL=0.689$, and $w=-1$.
\end{itemize}

All simulated events are analyzed as follows:
\begin{itemize}
    \setlength\itemsep{0.0em}
    \item Use SALT2 light-curve fit for each event to determine $\{t_0,m_B,x_1,c\}$.
    \item Apply selection requirements (cuts):
      \vspace{-0.2cm}
    \begin{itemize}
        \setlength\itemsep{0.01em}
        \item at least two passbands with maximum signal-to-noise ratio SNR$>5$;
        \item at least one \obs\ before $t_0$;
        \item at least one \obs\ $>10$~days after $t_0$ (rest frame);
        \item $|x_1| < 3$ and $|c|<0.3$;
        \item fitted uncertainties $\sigma_{x1}{<}1.0$ and $\sigma_{t0}{<}2$~days;
        \item SALT2 light curve fit probability $> 0.001$;
        \item valid bias correction in the BBC fit (see below).
    \end{itemize}
    For the \NSAMPLE\ samples, the average number of events passing cuts is 
    \NSNavgTOT\ (\NSNavgDES, \NSNavgFOUND, and \NSNavgLOWZ\ for 
    DES, Foundation, and \LOWZ, respectively).
    \item Determine $\PIa$ using the ``SuperNNova'' photometric classification \citep{Moller2020}\footnote{\URLSNN}
        based on recurrent neural networks.
    \item Use the BBC fit to determine redshift-binned HD corrected for 
           selection effects and non-SNIa contamination. We use $\nbz$ $z$-bins,
           with bin size proportional to $(1+z)^3$ so that there is finer $z$-binning
           at lower redshift.
    \item Create statistical+systematic covariance matrix as in \citet{Conley2011}.
    \item Use methods from \S\ref{sec:unbin} to produce an unbinned HD and two rebinned HDs.
      The first rebinned HD has \nbxA\ stretch and \nbcA\ color bins (\RebinA), and the total number of
      HD bins is $\nbz\times \nbxA\times \nbcA = \nbtotA$. 
      The second rebinned HD has \nbxB\ stretch and \nbcB\ color bins (\RebinB),
      and a total of \nbtotB\ HD bins. 
    \item Perform cosmology fit using a fast minimization program that combines an
      SN Ia HD with a cosmic microwave background (CMB) 
      prior that uses an $R$-shift parameter computed from the same 
      cosmology parameters as in the simulated samples.
      To match the constraining power from \citet{Planck2018}, the $R$-uncertainty is $\sigma_R=\sigR$.
      We fit for $\OM$ and $w$ using the \wCDM\ model, 
      and we also fit for {${\OM},w_0,w_a$} using the \wwCDM\ model.
  \item For the \wwCDM\ model, the figure of merit (FoM) is computed based on the 
        dark energy task force (DETF) definition in 
       \citet{Albrecht2006},
           \begin{equation}
               {\rm FoM}  = [ \sigma(w_0) \times \sigma(w_a) \times \sqrt{1-\rho^2} ]^{-1}
               \label{eq:FoM}
         \end{equation}
        where $\rho$ is the reduced covariance between $w_0$ and $w_a$.
\end{itemize}

Here we consider \NSYST\ systematic uncertainties that include the following:
\begin{itemize}
    \setlength\itemsep{0.01em}
    \item For each of the \NBAND\ passbands, shift the zero point using the uncertainty  
          from \citet{Frag2022},
    \item For each of the \NBAND\ passbands, shift the filter transmission wavelength   
          using the uncertainty from \citet{Frag2022},    
    \item Correlated zero point shift, $0.00714\lambda/{\rm micron}$,
        corresponding to the Hubble Space Telescope calibration uncertainty for primary reference C26202.
    \item Galactic extinction uncertainty is 4\%.
\end{itemize}
For each of the 68 zero-point and wavelength systematics, the SALT2 model is retrained and
the shift is propagated in the simulated data.
{\MVinprep} present the complete set of systematic uncertainties that includes
calibration covariances.

% =========================================================
\section{Validation II: Results}
\label{sec:validate2}

\subsection{BBC Sensitivity to Reference Cosmology}
\label{subsec:Cref}

\newcommand{\STDmu}{\rm{STD}_{\mu}}
\newcommand{\muiC}{\mu_{i,\C}} 
\newcommand{\muiCref}{\mu_{i,\Cref}} 

We begin by evaluating the sensitivity of fitted BBC distances to the 
reference cosmology $\Cref$ defined in Eq.~\ref{eq:HR}. 
Using the \wCDM\ model to vary $\Cref$, we vary $\OM$ up to $\pm 0.1$ with fixed $w=\wref$, 
and we vary $w$ up to $\pm 0.2$ with fixed $\OM=\OMref$. 
We define a sensitivity metric to be
\begin{equation}
    \STDmu = \rm{STD}(\muiC - \muiCref)
\end{equation}
where STD is the standard deviation,
$\muiCref$ are unbinned distances (Eq.~\ref{eq:mu}) using $\Cref$ (Eq.~\ref{eq:Cref}) 
in the BBC fit,
and $\muiC$ are unbinned distances from using a different $\C$ in the BBC fit.
Results for  8 \wCDM\ model variants are shown in Table~\ref{tb:STDmu} 
and we find $\STDmu \sim 10^{-4}$~mag,
which is about a 1000 times smaller than the intrinsic scatter.

The last 3 rows of Table~\ref{tb:STDmu} are based on a polynomial function of redshift for $\muiC$
to illustrate the BBC performance with poorer $\muiC$ estimates.
A constant $\muiC$ (p0) results in $\STDmu$ that is more than 1 order of magnitude larger
than for the \wCDM\ models, but is still well below 0.01~mag and thus works surprisingly well
for such a poor $\muiC$ estimate. Using 3rd- and 6th-order polynomial fits to the
baseline \LCDM\ model (p3 and p6) works much better than constant $\muiC$, 
but still not quite as well as for the \wCDM\ models.

\begin{table}[ht!]
\begin{center}
\caption{$\STDmu$ for Different $\C$ Choices}
\begin{tabular}{| l | c |}
\hline
$\C$ Variant & $\STDmu \times 10^4$ \\ \hline
 $\OM = 0.20$ & 1.0 \\
 $\OM = 0.25$ & 1.7 \\
 $\OM = 0.35$ & 0.4 \\
 $\OM = 0.40$ & 0.7 \\
 $w=-1.2$    & 0.7 \\
 $w=-1.1$    & 0.5 \\
 $w=-0.9$    & 0.3 \\
 $w=-0.8$    & 0.6 \\ \hline
 p0 ($\muiC=40$) & 59 \\
 p3\tablenote{3rd order polynomial fit to \LCDM\ model}      & 6.8 \\
 p6\tablenote{6th order polynomial fit to \LCDM\ model}      & 3.1  \\ \hline 
\end{tabular}
\label{tb:STDmu}
\end{center}
\vspace{-0.3cm}
\end{table}

% -- - - - - - - - - - - - - - - - - - - - - - - - - - 
\subsection{Uncertainty Scale and HR check}
\label{subsec:errscale}

\begin{figure}[hb]
    \includegraphics[width=0.9\linewidth]{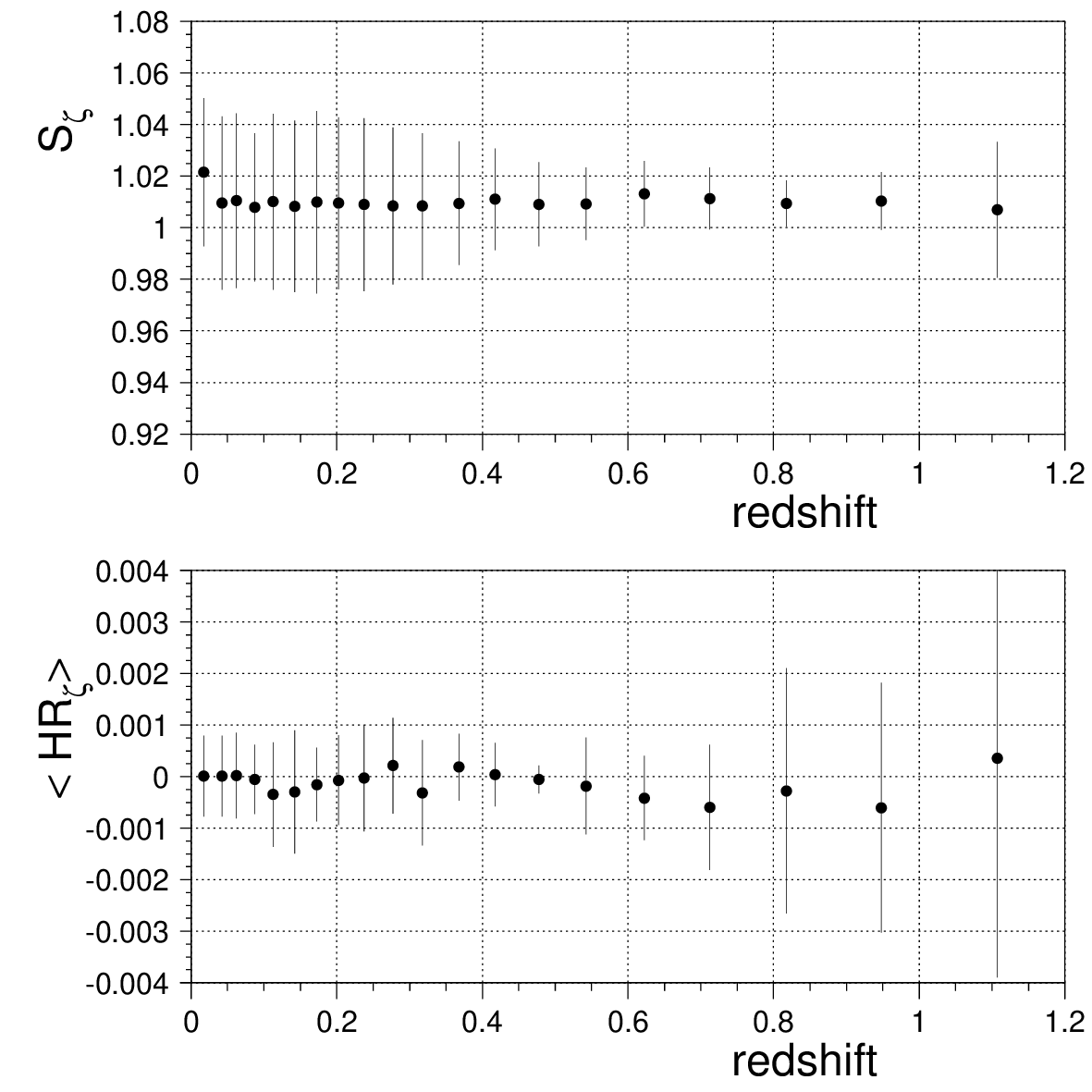}
    \caption{$\Sbin$ vs. redshift (upper panel) and 
    $\HRavg$ vs. redshift (lower panel). The error bars show
    the rms among the \NSAMPLE\ simulated data samples.
    }
    \label{fig:errscale}
\end{figure}

For the Ia+CC samples,
the uncertainty scale ($\Sbin$ in Eqs.~\ref{eq:sigM2} and \ref{eq:sigunbin}) 
vs. redshift bin is shown in the upper panel of Fig.~\ref{fig:errscale}, 
averaged over the \NSAMPLE\ data samples.
The average $\Sbin$ value is $\sim\Sbinavg$ at all redshifts, 
and is thus a small correction.
The Hubble residual crosscheck ($\HRavg$ defined in Eq.~\ref{eq:HRavg})
vs. redshift is shown in the lower panel of Fig.~\ref{fig:errscale}.
The values are within 0.001~mag of the expected value of zero. 

% - - - - - - - - - - - - - - - - - - - - - - -
\subsection{Bias Results for \wCDM}
\label{subsec:bias_wcdm}

We define the $w$-bias to be $\wbias \equiv w - w_{\rm true}$,
and we define \wbiasavg\ to be the average over the \NSAMPLE\ simulated data samples.
The \wbiasavg\ uncertainty is the standard deviation of the
$\wbias$ values divided by $\sqrt{\NSAMPLE}$.
We begin with an SN~Ia only sample that has no contamination and $\PIa=1$ for all events.
The \wbiasavg\ results are shown in the upper half of Table~\ref{tb:bias_wCDM} 
with no systematics (i.e., only statistical uncertainties),
with systematics included, and with four binning options: 
(i) binned, 
(ii) unbinned, 
(iii) rebinned with \nbxA\ stretch bins and \nbcA\ color bins (\RebinA), and
(iv) rebinned with \nbxB\ stretch bins and \nbcB\ color bins (\RebinB).
A bias consistent with zero at the $2\sigma$ level 
(\Nsigma\ in Table~\ref{tb:bias_wCDM})
is considered to be unbiased.
All \wbiasavg\ results are unbiased except for a $2.4\sigma$ bias for 
Ia-Only using an unbinned HD and systematics.
By averaging results over {\NSAMPLE} samples, 
the bias uncertainty and constraint is 
nearly 1 order of magnitude smaller than the 
average $w$-uncertainty (\sigwavg\ in Table~\ref{tb:bias_wCDM}) for a single data sample.

\newcommand{\wBiasDif}{$0.006\pm 0.001$}   %  wcdm: Rebin4x8 - unbin for Ia+CC, ALL syst

%In the lower half of Table~\ref{tb:bias_wCDM}, \wbiasavg\  is shown for the
%Ia+CC sample that includes contamination: all \wbiasavg\ results are unbiased.
%While each result is unbiased, the unbinned and \RebinB\ $w$-biases have
%opposite sign leading to a significant \wbiasavg\ difference of {\wBiasDif}.
%This difference is likely because a rebinned HD uses less information compared to the unbinned case.

The primary motivation for an unbinned HD is to reduce the total uncertainty.
With systematics, the average total uncertainty (\sigwavg\ column in Table~\ref{tb:bias_wCDM}) 
is reduced by ${\sim}7$\% compared to the binned result. 
% 309-250 = 182 ; 288-250 = 143
Naively subtracting the no-syst contribution in quadrature, the systematic uncertainties are 
0.0182 and 0.0143 for the binned and unbinned, respectively,
resulting in an ${\sim}20$\% reduction in the systematic uncertainty.
The rebinned uncertainties are comparable to that of the unbinned result.

Our 20\% reduction in systematic uncertainty is smaller than the 50\% 
reduction in BHS21 because we did not include the intrinsic scatter
systematic that is reduced by more than a factor of 3. 
For the calibration systematics used in both analyses, 
the reduction in BHS21 (see their Table 2) is similar to ours.

\begin{table*}
\begin{center}
\caption{  Average $w$-Bias, Significance, and Uncertainty vs. Redshift Binning Option ($w$CDM) }
\begin{tabular}{| l | l | l | lcc | } 
\hline 
 SN types &  Syst & Bin Option & \wbiasavg\ & \Nsigma\tablenote{\NsigwBiasFN} & \sigwavg\ \\ 
\hline 
\hline  
 Ia-Only  & None   & Binned       & $+0.0002 \pm 0.0030$  & $0.1$ & $0.0249$ \\ 
          &        & Unbin        & $+0.0004 \pm 0.0030$  & $0.1$ & $0.0250$ \\ 
          &        & \RebinA\tablenote{\RebinFNA} & $+0.0004 \pm 0.0030$  & $0.1$ & $0.0249$ \\ 
          &        & \RebinB\tablenote{\RebinFNB} & $+0.0005 \pm 0.0030$  & $0.2$ & $0.0250$ \\ 
\hline  
 Ia-Only  & All    & Binned       & $+0.0008 \pm 0.0031$  & $0.3$ & $0.0308$ \\ 
          &        & Unbin        & $-0.0069 \pm 0.0029$  & $2.4$ & $0.0288$ \\ 
          &        & \RebinA      & $+0.0009 \pm 0.0031$  & $0.3$ & $0.0293$ \\ 
          &        & \RebinB      & $+0.0010 \pm 0.0030$  & $0.3$ & $0.0286$ \\ 
\hline  
\hline  
 Ia+CC    & None   & Binned       & $+0.0020 \pm 0.0030$  & $0.7$ & $0.0250$ \\ 
          &        & Unbin        & $+0.0026 \pm 0.0030$  & $0.9$ & $0.0250$ \\ 
          &        & \RebinA      & $+0.0022 \pm 0.0030$  & $0.7$ & $0.0250$ \\ 
          &        & \RebinB      & $+0.0022 \pm 0.0030$  & $0.7$ & $0.0250$ \\ 
\hline  
 Ia+CC    & All    & Binned       & $+0.0024 \pm 0.0032$  & $0.7$ & $0.0309$ \\ 
          &        & Unbin        & $-0.0044 \pm 0.0030$  & $1.5$ & $0.0288$ \\ 
          &        & \RebinA      & $+0.0025 \pm 0.0031$  & $0.8$ & $0.0294$ \\ 
          &        & \RebinB      & $+0.0019 \pm 0.0032$  & $0.6$ & $0.0286$ \\ 
\hline  
\end{tabular} 
\label{tb:bias_wCDM} 
\end{center} 
\vspace{-0.3cm} 
\end{table*}

% - - - - - - - - - - - - - - - - - - - - 
\subsection{Bias Results for \wwCDM}
\label{subsec:bias_wwcdm}

\newcommand{\woBiasDif}{$-0.013 \pm 0.006$} % Rebin4x8 - unbin for Ia+CC, ALL syst
\newcommand{\waBiasDif}{$0.09 \pm 0.03$}

For the \wwCDM\ model, the bias summary is shown in Table~\ref{tb:bias_w0waCDM}.
The biases are consistent with zero at the $2\sigma$ level,
and the bias uncertainty and constraint are nearly an order of magnitude smaller
than the average single-sample uncertainty 
(\sigwoavg\ and \sigwaavg\ in Table~\ref{tb:bias_w0waCDM}).
For the Ia+CC sample, the average FoM is {\FoMavg}$=\FoMbin$ for the binned option,
and the unbinned option increases \FoMavg\ to $\FoMunbin$.
The \RebinA\ option results in {\FoMavg}$=\FoMrebinA$ that is between the 
binned and unbinned \FoMavg.
The \RebinB\ option results in {\FoMavg}$=\FoMrebinB$ that is very close to the unbinned \FoMavg.

%As with the \wCDM\ case, each Ia+CC binning option is unbiased but there is evidence
%for a difference between unbinned and \RebinB\ bias: 
%{\woBiasDif} for {\wobiasavg} difference and {\waBiasDif} for \wabiasavg\ difference.

\begin{table*}
\begin{center}
\caption{  Average $w0,w_a$-bias, Significance, Uncertainty, and FoM vs. Redshift Binning Option ($w_0w_a$CDM) }
\begin{tabular}{| l | l | l | lcclccc | } 
\hline 
 SN types &  Syst & Bin Option & \wobiasavg\ & \Nsigma\tablenote{\NsigwoBiasFN} & \sigwoavg\ &\wabiasavg\ & \Nsigma\tablenote{\NsigwaBiasFN} & \sigwaavg\ & \FoMavg\ \\ 
\hline 
\hline  
 Ia-Only  & None   & Binned       & $-0.005 \pm 0.014$  & $0.3$ & $0.101$ & $-0.022 \pm 0.063$  & $0.3$ & $0.482$ & 78 \\ 
          &        & Unbin        & $-0.004 \pm 0.014$  & $0.3$ & $0.100$ & $-0.023 \pm 0.063$  & $0.4$ & $0.480$ & 78 \\ 
          &        & \RebinA\tablenote{\RebinFNA} & $-0.005 \pm 0.014$  & $0.4$ & $0.101$ & $-0.017 \pm 0.064$  & $0.3$ & $0.482$ & 78 \\ 
          &        & \RebinB\tablenote{\RebinFNB} & $-0.006 \pm 0.014$  & $0.4$ & $0.101$ & $-0.015 \pm 0.064$  & $0.2$ & $0.481$ & 78 \\ 
\hline  
 Ia-Only  & All    & Binned       & $-0.003 \pm 0.015$  & $0.2$ & $0.138$ & $-0.049 \pm 0.069$  & $0.7$ & $0.634$ & 45 \\ 
          &        & Unbin        & $+0.008 \pm 0.014$  & $0.5$ & $0.122$ & $-0.122 \pm 0.066$  & $1.9$ & $0.575$ & 56 \\ 
          &        & \RebinA      & $-0.002 \pm 0.015$  & $0.1$ & $0.130$ & $-0.047 \pm 0.069$  & $0.7$ & $0.595$ & 51 \\ 
          &        & \RebinB      & $-0.001 \pm 0.014$  & $0.1$ & $0.125$ & $-0.044 \pm 0.065$  & $0.7$ & $0.576$ & 55 \\ 
\hline  
\hline  
 Ia+CC    & None   & Binned       & $-0.010 \pm 0.014$  & $0.7$ & $0.102$ & $+0.013 \pm 0.065$  & $0.2$ & $0.481$ & 77 \\ 
          &        & Unbin        & $-0.010 \pm 0.014$  & $0.7$ & $0.101$ & $+0.013 \pm 0.065$  & $0.2$ & $0.479$ & 78 \\ 
          &        & \RebinA      & $-0.013 \pm 0.014$  & $0.9$ & $0.102$ & $+0.025 \pm 0.066$  & $0.4$ & $0.481$ & 77 \\ 
          &        & \RebinB      & $-0.013 \pm 0.014$  & $0.9$ & $0.102$ & $+0.025 \pm 0.065$  & $0.4$ & $0.480$ & 77 \\ 
\hline  
 Ia+CC    & All    & Binned       & $-0.004 \pm 0.015$  & $0.3$ & $0.139$ & $-0.039 \pm 0.069$  & $0.6$ & $0.631$ & 45 \\ 
          &        & Unbin        & $+0.005 \pm 0.015$  & $0.3$ & $0.123$ & $-0.100 \pm 0.067$  & $1.5$ & $0.574$ & 55 \\ 
          &        & \RebinA      & $-0.003 \pm 0.016$  & $0.2$ & $0.131$ & $-0.037 \pm 0.072$  & $0.5$ & $0.595$ & 51 \\ 
          &        & \RebinB      & $-0.008 \pm 0.015$  & $0.5$ & $0.126$ & $-0.012 \pm 0.069$  & $0.2$ & $0.575$ & 54 \\ 
\hline  
\end{tabular} 
\label{tb:bias_w0waCDM} 
\end{center} 
\vspace{-0.3cm} 
\end{table*}

\subsection{Binning Option Consistency} 
\label{subsec:compare}

Without systematics, \wCDM\ $w$-results for all binning options,
with and without contamination, agree to within $<0.001$.
For the \wwCDM\ model and Ia-Only, all binning options agree to within
$0.001$ and $<0.01$ for $w_0$ and $w_a$, respectively. With contamination,
the rebin results differ by $0.003$ and $0.01$ for $w_0$ and $w_a$, respectively,
suggesting a subtle bias with the rebin procedure.

With systematics, the binned and rebinned \wCDM\ results agree to within $0.001$ in $w$,
while the unbinned $w$-result differs significantly by $0.006\pm 0.001$,
which corresponds to 20\% of the total uncertainty.
This comparison is the same with and without contamination.
For the \wwCDM\ model, the binned and rebinned results agree to within $0.002$ in $w_0$;
the $w_a$ results agree to within $<0.01$ for Ia-Only and differ by up to nearly 0.03
with contamination. The unbinned results differ by ${\sim}0.01$ and $0.06$
for $w_0$ and $w_a$, respectively.
While all binning options show unbiased cosmology results,
there is evidence for a small difference between the binned the unbinned results.
This difference is present with our without contamination.

% - - - - - - - - - - - - - - - - - - -  
\subsection{Impact of $\PBIa$ Term} 
\label{subsec:bias_noPIa}

To check the impact of the $\PBIa$ term in Eq.~\ref{eq:sigunbin},
we forced $\PBIa=1$ and reevaluated the \wwCDM\ bias for an unbinned HD.
We find more than $10\sigma$ bias, which illustrates the necessity
of accurately evaluating this term.

% ================================
\section{Conclusion}
As a follow up to the original binned Hubble diagram (HD) from BBC,
we have developed methods to derive cosmological results from an unbinned HD,
and from a rebinned HD in the space of redshift, stretch, and color.
Averaging analysis results from \NSAMPLE\ simulated data samples,
we find biases consistent with zero and bias constraints almost 1 order
of magnitude smaller than the single-sample uncertainty. We also find
that using an unbinned HD results in a reduced total uncertainty consistent
with BHS21.
This conclusion holds for both the \wCDM\ and \wwCDM\ models, 
and we find the same results with or without photometric contamination.

Using a rebinned HD with \nbxA\ stretch and \nbcA\ color bins (\RebinA), 
we recover unbiased cosmology results and also benefit from the 
reduced uncertainty in the unbinned HD.
Using more bins (\nbxB\ stretch and \nbcB\ color bins), 
there is still no bias and the total uncertainty is similar to the unbinned case.
With ${\sim}2000$ events in the DES unbinned HD, the rebinned cosmology-fitting speed is only
a factor of few faster compared to the unbinned case. 
With anticipated future samples in the $10^4$-$10^5$ range,
the rebinned HD size does not increase and therefore the
cosmology-fitting speed improvement will be much more significant.

While our unbiased results are encouraging, we note that
\cite{Mitra2023} reported a significant cosmology bias
using an unbinned HD from a simulated
LSST data sample of pure SNe~Ia. We therefore recommend
repeating our bias tests on simulated data for future analyses.

% ================================
\section{Acknowledgements}

R.K. is supported by DOE grant DE-SC0009924.
P.A acknowledges parts of this research were carried out on the traditional lands of the 
Ngunnawal people.  We pay our respects to their elders past, present, and emerging.
P.A. was supported by an Australian Government Research Training Program (RTP) Scholarship.
We acknowledge the University of Chicago's 
Research Computing Center for their support of this work.

\clearpage
\bibliographystyle{apj}
\bibliography{main}

\begin{thebibliography}{}
\expandafter\ifx\csname natexlab\endcsname\relax\def\natexlab#1{#1}\fi

\bibitem[{{Albrecht} {et~al.}(2006){Albrecht}, {Bernstein}, {Cahn}, {Freedman},
  {Hewitt}, {Hu}, {Huth}, {Kamionkowski}, {Kolb}, {Knox}, {Mather}, {Staggs},
  \& {Suntzeff}}]{Albrecht2006}
{Albrecht}, A., {Bernstein}, G., {Cahn}, R., {et~al.} 2006, arXiv e-prints,
  astro

\bibitem[{{Betoule} {et~al.}(2014){Betoule}, {Kessler}, {Guy}, {Mosher},
  {Hardin}, {Biswas}, {Astier}, {El-Hage}, {Konig}, {Kuhlmann}, {Marriner},
  {Pain}, {Regnault}, {Balland}, {Bassett}, {Brown}, {Campbell}, {Carlberg},
  {Cellier-Holzem}, {Cinabro}, {Conley}, {D'Andrea}, {DePoy}, {Doi}, {Ellis},
  {Fabbro}, {Filippenko}, {Foley}, {Frieman}, {Fouchez}, {Galbany}, {Goobar},
  {Gupta}, {Hill}, {Hlozek}, {Hogan}, {Hook}, {Howell}, {Jha}, {Le Guillou},
  {Leloudas}, {Lidman}, {Marshall}, {M{\"o}ller}, {Mour{\~a}o}, {Neveu},
  {Nichol}, {Olmstead}, {Palanque-Delabrouille}, {Perlmutter}, {Prieto},
  {Pritchet}, {Richmond}, {Riess}, {Ruhlmann-Kleider}, {Sako}, {Schahmaneche},
  {Schneider}, {Smith}, {Sollerman}, {Sullivan}, {Walton}, \&
  {Wheeler}}]{Betoule14}
{Betoule}, M., {Kessler}, R., {Guy}, J., {et~al.} 2014, \aap, 568, A22

\bibitem[{{Brout} {et~al.}(2021){Brout}, {Hinton}, \&
  {Scolnic}}]{Binning_is_Sinning}
{Brout}, D., {Hinton}, S.~R., \& {Scolnic}, D. 2021, \apjl, 912, L26

\bibitem[{{Brout} {et~al.}(2022{\natexlab{a}}){Brout}, {Scolnic}, {Popovic},
  {Riess}, {Carr}, {Zuntz}, {Kessler}, {Davis}, {Hinton}, {Jones}, {Kenworthy},
  {Peterson}, {Said}, {Taylor}, {Ali}, {Armstrong}, {Charvu}, {Dwomoh},
  {Meldorf}, {Palmese}, {Qu}, {Rose}, {Sanchez}, {Stubbs}, {Vincenzi}, {Wood},
  {Brown}, {Chen}, {Chambers}, {Coulter}, {Dai}, {Dimitriadis}, {Filippenko},
  {Foley}, {Jha}, {Kelsey}, {Kirshner}, {M{\"o}ller}, {Muir}, {Nadathur},
  {Pan}, {Rest}, {Rojas-Bravo}, {Sako}, {Siebert}, {Smith}, {Stahl}, \&
  {Wiseman}}]{PantheonPlus}
{Brout}, D., {Scolnic}, D., {Popovic}, B., {et~al.} 2022{\natexlab{a}}, \apj,
  938, 110

\bibitem[{{Brout} {et~al.}(2022{\natexlab{b}}){Brout}, {Taylor}, {Scolnic},
  {Wood}, {Rose}, {Vincenzi}, {Dwomoh}, {Lidman}, {Riess}, {Ali}, {Qu}, \&
  {Dai}}]{Frag2022}
{Brout}, D., {Taylor}, G., {Scolnic}, D., {et~al.} 2022{\natexlab{b}}, \apj,
  938, 111

\bibitem[{{Conley} {et~al.}(2011){Conley}, {Guy}, {Sullivan}, {Regnault},
  {Astier}, {Balland}, {Basa}, {Carlberg}, {Fouchez}, {Hardin}, {Hook},
  {Howell}, {Pain}, {Palanque-Delabrouille}, {Perrett}, {Pritchet}, {Rich},
  {Ruhlmann-Kleider}, {Balam}, {Baumont}, {Ellis}, {Fabbro}, {Fakhouri},
  {Fourmanoit}, {Gonz{\'a}lez-Gait{\'a}n}, {Graham}, {Hudson}, {Hsiao},
  {Kronborg}, {Lidman}, {Mourao}, {Neill}, {Perlmutter}, {Ripoche}, {Suzuki},
  \& {Walker}}]{Conley2011}
{Conley}, A., {Guy}, J., {Sullivan}, M., {et~al.} 2011, \apjs, 192, 1

\bibitem[{{DES~Collaboration}(2019)}]{des3yr}
{DES~Collaboration}. 2019, \apjl, 872, L30

\bibitem[{{Faccioli} {et~al.}(2011){Faccioli}, {Kim}, {Miquel}, {Bernstein},
  {Bonissent}, {Brown}, {Carithers}, {Christiansen}, {Connolly}, {Deustua},
  {Gerdes}, {Gladney}, {Kushner}, {Linder}, {McKee}, {Mostek}, {Shukla},
  {Stebbins}, {Stoughton}, \& {Tucker}}]{SelfCalib2011}
{Faccioli}, L., {Kim}, A.~G., {Miquel}, R., {et~al.} 2011, Astroparticle
  Physics, 34, 847

\bibitem[{{Foley} {et~al.}(2018){Foley}, {Scolnic}, {Rest}, {Jha}, {Pan},
  {Riess}, {Challis}, {Chambers}, {Coulter}, {Dettman}, {Foley}, {Fox},
  {Huber}, {Jones}, {Kilpatrick}, {Kirshner}, {Schultz}, {Siebert},
  {Flewelling}, {Gibson}, {Magnier}, {Miller}, {Primak}, {Smartt}, {Smith},
  {Wainscoat}, {Waters}, \& {Willman}}]{Found2018}
{Foley}, R.~J., {Scolnic}, D., {Rest}, A., {et~al.} 2018, \mnras, 475, 193

\bibitem[{{Guy} {et~al.}(2010){Guy}, {Sullivan}, {Conley}, {Regnault},
  {Astier}, {Balland}, {Basa}, {Carlberg}, {Fouchez}, {Hardin}, {Hook},
  {Howell}, {Pain}, {Palanque-Delabrouille}, {Perrett}, {Pritchet}, {Rich},
  {Ruhlmann-Kleider}, {Balam}, {Baumont}, {Ellis}, {Fabbro}, {Fakhouri},
  {Fourmanoit}, {Gonz{\'a}lez-Gait{\'a}n}, {Graham}, {Hsiao}, {Kronborg},
  {Lidman}, {Mourao}, {Perlmutter}, {Ripoche}, {Suzuki}, \& {Walker}}]{Guy2010}
{Guy}, J., {Sullivan}, M., {Conley}, A., {et~al.} 2010, \aap, 523, A7

\bibitem[{{Hicken} {et~al.}(2009){Hicken}, {Challis}, {Jha}, {Kirshner},
  {Matheson}, {Modjaz}, {Rest}, {Wood-Vasey}, {Bakos}, {Barton}, {Berlind},
  {Bragg}, {Brice{\~n}o}, {Brown}, {Caldwell}, {Calkins}, {Cho}, {Ciupik},
  {Contreras}, {Dendy}, {Dosaj}, {Durham}, {Eriksen}, {Esquerdo}, {Everett},
  {Falco}, {Fernandez}, {Gaba}, {Garnavich}, {Graves}, {Green}, {Groner},
  {Hergenrother}, {Holman}, {Hradecky}, {Huchra}, {Hutchison}, {Jerius},
  {Jordan}, {Kilgard}, {Krauss}, {Luhman}, {Macri}, {Marrone}, {McDowell},
  {McIntosh}, {McNamara}, {Megeath}, {Mochejska}, {Munoz}, {Muzerolle},
  {Naranjo}, {Narayan}, {Pahre}, {Peters}, {Peterson}, {Rines}, {Ripman},
  {Roussanova}, {Schild}, {Sicilia-Aguilar}, {Sokoloski}, {Smalley}, {Smith},
  {Spahr}, {Stanek}, {Barmby}, {Blondin}, {Stubbs}, {Szentgyorgyi}, {Torres},
  {Vaz}, {Vikhlinin}, {Wang}, {Westover}, {Woods}, \& {Zhao}}]{Hicken2009_CfA3}
{Hicken}, M., {Challis}, P., {Jha}, S., {et~al.} 2009, \apj, 700, 331

\bibitem[{{Hicken} {et~al.}(2012){Hicken}, {Challis}, {Kirshner}, {Rest},
  {Cramer}, {Wood-Vasey}, {Bakos}, {Berlind}, {Brown}, {Caldwell}, {Calkins},
  {Currie}, {de Kleer}, {Esquerdo}, {Everett}, {Falco}, {Fernandez},
  {Friedman}, {Groner}, {Hartman}, {Holman}, {Hutchins}, {Keys}, {Kipping},
  {Latham}, {Marion}, {Narayan}, {Pahre}, {Pal}, {Peters}, {Perumpilly},
  {Ripman}, {Sipocz}, {Szentgyorgyi}, {Tang}, {Torres}, {Vaz}, {Wolk}, \&
  {Zezas}}]{Hicken2012}
{Hicken}, M., {Challis}, P., {Kirshner}, R.~P., {et~al.} 2012, \apjs, 200, 12

\bibitem[{{Hinton} \& {Brout}(2020)}]{Pippin}
{Hinton}, S., \& {Brout}, D. 2020, JOSS, 5, 2122

\bibitem[{{Hlozek} {et~al.}(2012){Hlozek}, {Kunz}, {Bassett}, {Smith},
  {Newling}, {et~al.}}]{BEAMS2012}
{Hlozek}, R., {Kunz}, M., {Bassett}, B., {et~al.} 2012, \apj, 752, 79

\bibitem[{{Jones} {et~al.}(2019){Jones}, {Scolnic}, {Foley}, {Rest}, {Kessler},
  {et~al.}}]{Jones2019}
{Jones}, D.~O., {Scolnic}, D.~M., {Foley}, R.~J., {et~al.} 2019, \apj, 881, 19

\bibitem[{{Jones} {et~al.}(2018){Jones}, {Scolnic}, {Riess}, {Rest},
  {Kirshner}, {et~al.}}]{Jones2018}
{Jones}, D.~O., {Scolnic}, D.~M., {Riess}, A.~G., {et~al.} 2018, \apj, 857, 51

\bibitem[{{Kessler} {et~al.}(2019{\natexlab{a}}){Kessler}, {Brout}, {Crawford},
  {et~al.}}]{Kessler2019_sim}
{Kessler}, R., {Brout}, D., {Crawford}, S., {et~al.} 2019{\natexlab{a}},
  \mnras, 485, 1171

\bibitem[{{Kessler} \& {Scolnic}(2017)}]{bbc}
{Kessler}, R., \& {Scolnic}, D. 2017, \apj, 836, 56

\bibitem[{{Kessler} {et~al.}(2009){Kessler}, {Bernstein}, {Cinabro}, {Dilday},
  {Frieman}, {Jha}, {Kuhlmann}, {Miknaitis}, {Sako}, {Taylor}, \&
  {Vanderplas}}]{Kessler09SNANA}
{Kessler}, R., {Bernstein}, J.~P., {Cinabro}, D., {et~al.} 2009, \pasp, 121,
  1028

\bibitem[{{Kessler} {et~al.}(2019{\natexlab{b}}){Kessler}, {Narayan},
  {Avelino}, {Bachelet}, {Biswas}, {Brown}, {Chernoff}, {Connolly}, {Dai},
  {Daniel}, {Di Stefano}, {Drout}, {Galbany}, {Gonz{\'a}lez-Gait{\'a}n},
  {Graham}, {Hlo{\v{z}}ek}, {Ishida}, {Guillochon}, {Jha}, {Jones}, {Mandel},
  {Muthukrishna}, {O'Grady}, {Peters}, {Pierel}, {Ponder}, {Pr{\v{s}}a},
  {Rodney}, {Villar}, {LSST Dark Energy Science Collaboration}, \& {Transient
  and Variable Stars Science Collaboration}}]{PLASTICC}
{Kessler}, R., {Narayan}, G., {Avelino}, A., {et~al.} 2019{\natexlab{b}},
  \pasp, 131, 094501

\bibitem[{{Kunz} {et~al.}(2007){Kunz}, {Bassett}, \& {Hlozek}}]{BEAMS2007}
{Kunz}, M., {Bassett}, B.~A., \& {Hlozek}, R.~A. 2007, \prd, 75, 103508

\bibitem[{{Lochner} {et~al.}(2016){Lochner}, {McEwen}, {Peiris}, {Lahav}, \&
  {Winter}}]{Lochner2016}
{Lochner}, M., {McEwen}, J.~D., {Peiris}, H.~V., {Lahav}, O., \& {Winter},
  M.~K. 2016, \apjs, 225, 31

\bibitem[{{Marriner} {et~al.}(2011){Marriner}, {Bernstein}, {Kessler},
  {Lampeitl}, {Miquel}, {et~al.}}]{Marriner2011}
{Marriner}, J., {Bernstein}, J.~P., {Kessler}, R., {et~al.} 2011, \apj, 740, 72

\bibitem[{{Mitra} {et~al.}(2023){Mitra}, {Kessler}, {More}, {Hlozek}, \& {LSST
  Dark Energy Science Collaboration}}]{Mitra2023}
{Mitra}, A., {Kessler}, R., {More}, S., {Hlozek}, R., \& {LSST Dark Energy
  Science Collaboration}. 2023, \apj, 944, 212

\bibitem[{{M{\"o}ller} \& {de Boissi{\`e}re}(2020)}]{Moller2020}
{M{\"o}ller}, A., \& {de Boissi{\`e}re}, T. 2020, \mnras, 491, 4277

\bibitem[{{Perlmutter} {et~al.}(1999){Perlmutter}, {Aldering}, {Goldhaber},
  {Knop}, {Nugent}, {Castro}, {Deustua}, {Fabbro}, {Goobar}, {Groom}, {Hook},
  {Kim}, {Kim}, {Lee}, {Nunes}, {Pain}, {Pennypacker}, {Quimby}, {Lidman},
  {Ellis}, {Irwin}, {McMahon}, {Ruiz-Lapuente}, {Walton}, {Schaefer}, {Boyle},
  {Filippenko}, {Matheson}, {Fruchter}, {Panagia}, {Newberg}, {Couch}, \&
  {Project}}]{Perlmutter99}
{Perlmutter}, S., {Aldering}, G., {Goldhaber}, G., {et~al.} 1999, \apj, 517,
  565

\bibitem[{{Planck Collaboration} {et~al.}(2020){Planck Collaboration},
  {Aghanim}, {Akrami}, {Ashdown}, {Aumont}, {Baccigalupi}, {Ballardini},
  {Banday}, {Barreiro}, {Bartolo}, {Basak}, {Battye}, {Benabed}, {Bernard},
  {Bersanelli}, {Bielewicz}, {Bock}, {Bond}, {Borrill}, {Bouchet}, {Boulanger},
  {Bucher}, {Burigana}, {Butler}, {Calabrese}, {Cardoso}, {Carron},
  {Challinor}, {Chiang}, {Chluba}, {Colombo}, {Combet}, {Contreras}, {Crill},
  {Cuttaia}, {de Bernardis}, {de Zotti}, {Delabrouille}, {Delouis}, {Di
  Valentino}, {Diego}, {Dor{\'e}}, {Douspis}, {Ducout}, {Dupac}, {Dusini},
  {Efstathiou}, {Elsner}, {En{\ss}lin}, {Eriksen}, {Fantaye}, {Farhang},
  {Fergusson}, {Fernandez-Cobos}, {Finelli}, {Forastieri}, {Frailis},
  {Fraisse}, {Franceschi}, {Frolov}, {Galeotta}, {Galli}, {Ganga},
  {G{\'e}nova-Santos}, {Gerbino}, {Ghosh}, {Gonz{\'a}lez-Nuevo}, {G{\'o}rski},
  {Gratton}, {Gruppuso}, {Gudmundsson}, {Hamann}, {Handley}, {Hansen},
  {Herranz}, {Hildebrandt}, {Hivon}, {Huang}, {Jaffe}, {Jones}, {Karakci},
  {Keih{\"a}nen}, {Keskitalo}, {Kiiveri}, {Kim}, {Kisner}, {Knox},
  {Krachmalnicoff}, {Kunz}, {Kurki-Suonio}, {Lagache}, {Lamarre}, {Lasenby},
  {Lattanzi}, {Lawrence}, {Le Jeune}, {Lemos}, {Lesgourgues}, {Levrier},
  {Lewis}, {Liguori}, {Lilje}, {Lilley}, {Lindholm}, {L{\'o}pez-Caniego},
  {Lubin}, {Ma}, {Mac{\'\i}as-P{\'e}rez}, {Maggio}, {Maino}, {Mandolesi},
  {Mangilli}, {Marcos-Caballero}, {Maris}, {Martin}, {Martinelli},
  {Mart{\'\i}nez-Gonz{\'a}lez}, {Matarrese}, {Mauri}, {McEwen}, {Meinhold},
  {Melchiorri}, {Mennella}, {Migliaccio}, {Millea}, {Mitra},
  {Miville-Desch{\^e}nes}, {Molinari}, {Montier}, {Morgante}, {Moss}, {Natoli},
  {N{\o}rgaard-Nielsen}, {Pagano}, {Paoletti}, {Partridge}, {Patanchon},
  {Peiris}, {Perrotta}, {Pettorino}, {Piacentini}, {Polastri}, {Polenta},
  {Puget}, {Rachen}, {Reinecke}, {Remazeilles}, {Renzi}, {Rocha}, {Rosset},
  {Roudier}, {Rubi{\~n}o-Mart{\'\i}n}, {Ruiz-Granados}, {Salvati}, {Sandri},
  {Savelainen}, {Scott}, {Shellard}, {Sirignano}, {Sirri}, {Spencer},
  {Sunyaev}, {Suur-Uski}, {Tauber}, {Tavagnacco}, {Tenti}, {Toffolatti},
  {Tomasi}, {Trombetti}, {Valenziano}, {Valiviita}, {Van Tent}, {Vibert},
  {Vielva}, {Villa}, {Vittorio}, {Wandelt}, {Wehus}, {White}, {White},
  {Zacchei}, \& {Zonca}}]{Planck2018}
{Planck Collaboration}, {Aghanim}, N., {Akrami}, Y., {et~al.} 2020, \aap, 641,
  A6

\bibitem[{{Qu} {et~al.}(2021){Qu}, {Sako}, {M{\"o}ller}, \& {Doux}}]{Qu2021}
{Qu}, H., {Sako}, M., {M{\"o}ller}, A., \& {Doux}, C. 2021, \aj, 162, 67

\bibitem[{{Riess} {et~al.}(1998){Riess}, {Filippenko}, {Challis},
  {Clocchiatti}, {Diercks}, {Garnavich}, {Gilliland}, {Hogan}, {Jha},
  {Kirshner}, {Leibundgut}, {Phillips}, {Reiss}, {Schmidt}, {Schommer},
  {Smith}, {Spyromilio}, {Stubbs}, {Suntzeff}, \& {Tonry}}]{Riess98}
{Riess}, A.~G., {Filippenko}, A.~V., {Challis}, P., {et~al.} 1998, \aj, 116,
  1009

\bibitem[{{Sako} {et~al.}(2011){Sako}, {Bassett}, {Connolly}, {Dilday},
  {Cambell}, {Frieman}, {Gladney}, {Kessler}, {Lampeitl}, {Marriner}, {Miquel},
  {Nichol}, {Schneider}, {Smith}, \& {Sollerman}}]{Sako2011}
{Sako}, M., {Bassett}, B., {Connolly}, B., {et~al.} 2011, \apj, 738, 162

\bibitem[{{Scolnic} {et~al.}(2018){Scolnic}, {Jones}, {Rest}, {Pan},
  {Chornock}, {Foley}, {Huber}, {Kessler}, {Narayan}, {Riess}, {Rodney},
  {Berger}, {Brout}, {Challis}, {Drout}, {Finkbeiner}, {Lunnan}, {Kirshner},
  {Sanders}, {Schlafly}, {Smartt}, {Stubbs}, {Tonry}, {Wood-Vasey}, {Foley},
  {Hand}, {Johnson}, {Burgett}, {Chambers}, {Draper}, {Hodapp}, {Kaiser},
  {Kudritzki}, {Magnier}, {Metcalfe}, {Bresolin}, {Gall}, {Kotak}, {McCrum}, \&
  {Smith}}]{Pantheon}
{Scolnic}, D.~M., {Jones}, D.~O., {Rest}, A., {et~al.} 2018, \apj, 859, 101

\bibitem[{{Tripp}(1998)}]{Tripp1998}
{Tripp}, R. 1998, \aap, 331, 815

\bibitem[{{Vincenzi} {et~al.}(2019){Vincenzi}, {Sullivan}, {Firth},
  {Guti{\'e}rrez}, {Frohmaier}, {Smith}, {Angus}, \& {Nichol}}]{Vinc2019}
{Vincenzi}, M., {Sullivan}, M., {Firth}, R.~E., {et~al.} 2019, \mnras, 489,
  5802

\bibitem[{{Vincenzi} {et~al.}(2021){Vincenzi}, {Sullivan}, {Graur}, {Brout},
  {Davis}, {et~al.}}]{Vinc2021_sim}
{Vincenzi}, M., {Sullivan}, M., {Graur}, O., {et~al.} 2021, \mnras, 505, 2819

\bibitem[{{Vincenzi} {et~al.}(2023){Vincenzi}, {Sullivan}, {M{\"o}ller},
  {Armstrong}, {Bassett}, {Brout}, {Carollo}, {Carr}, {Davis}, {Frohmaier},
  {Galbany}, {Glazebrook}, {Graur}, {Kelsey}, {Kessler}, {Kovacs}, {Lewis},
  {Lidman}, {Malik}, {Nichol}, {Popovic}, {Sako}, {Scolnic}, {Smith}, {Taylor},
  {Tucker}, {Wiseman}, {Aguena}, {Allam}, {Annis}, {Asorey}, {Bacon}, {Bertin},
  {Brooks}, {Burke}, {Carnero Rosell}, {Carretero}, {Castander}, {Costanzi},
  {da Costa}, {Pereira}, {De Vicente}, {Desai}, {Diehl}, {Doel}, {Everett},
  {Ferrero}, {Flaugher}, {Fosalba}, {Frieman}, {Garc{\'\i}a-Bellido}, {Gerdes},
  {Gruen}, {Gutierrez}, {Hinton}, {Hollowood}, {Honscheid}, {James}, {Kuehn},
  {Kuropatkin}, {Lahav}, {Li}, {Lima}, {Maia}, {Marshall}, {Miquel}, {Morgan},
  {Ogando}, {Palmese}, {Paz-Chinch{\'o}n}, {Pieres}, {Plazas Malag{\'o}n},
  {Reil}, {Roodman}, {Sanchez}, {Schubnell}, {Serrano}, {Sevilla-Noarbe},
  {Suchyta}, {Tarle}, {To}, {Varga}, {Weller}, {Wilkinson}, \& {DES
  Collaboration}}]{Vinc2023_bias}
{Vincenzi}, M., {Sullivan}, M., {M{\"o}ller}, A., {et~al.} 2023, \mnras, 518,
  1106

\end{thebibliography}

\clearpage

\end{document}